\documentclass[11pt,a4paper]{article}
\pdfoutput=1
\usepackage{microtype}
\DisableLigatures[f]{encoding = *, family = * }
\usepackage{upref,amsmath,amssymb,amsthm,mathrsfs,epsf,amsfonts}
\usepackage{graphicx}
\usepackage{hyperref}
\usepackage{paralist}
\usepackage{multirow}
\usepackage{float}
\usepackage{url}
\usepackage{rotating}
\usepackage{epstopdf}
\usepackage{tablefootnote}
\usepackage{xfrac}
\usepackage{paralist}
\usepackage[round]{natbib}
\usepackage{caption}
\usepackage{subcaption}
\usepackage[table]{xcolor}
\usepackage[version=0.96]{pgf}
\usepackage{tikz}
\usetikzlibrary{arrows,shapes,snakes,automata,backgrounds,petri}
\usepackage{pgfplots}
\usepackage{authblk}
\usepackage[section]{placeins}
\usepackage{setspace}
\usepackage{diagbox}
\usepackage{enumitem}

\newlist{Description}{description}{3}
\setlist[Description]{style=nextline,itemsep=0mm}
\SetEnumitemKey{margin}{leftmargin={\widthof{#1}+2em}}

\graphicspath{{./figs/}}

\begin{document}
\title{Bayesian reconstruction of past land-cover from pollen data: model robustness
  and sensitivity to auxiliary variables}
\author[1,2,3]{Behnaz Pirzamanbein \thanks{Corresponding author: Behnaz Pirzamanbein, bepi@dtu.dk}}
\affil[1]{Department of Applied Mathematics and Computer Science, Technical University of Denmark}
\affil[2]{Centre for Mathematical Sciences, Lund University, Sweden}
\affil[3]{Centre for Environmental and Climate Research, Lund University, Sweden}
\author[1]{Johan Lindstr\"{o}m}
\author[4,5]{Anneli Poska}
\affil[4]{Department of Physical Geography and Ecosystems Analysis, Lund University, Sweden}
\affil[5]{Institute of Geology, Tallinn University of Technology, Estonia}
\renewcommand\Authands{ and }
\date{\vspace{-5ex}}
\maketitle

\begin{abstract}
Realistic depictions of past land cover are needed to investigate prehistoric
environmental changes, effects of anthropogenic deforestation, and long term
land cover-climate feedbacks. Observation based reconstructions of
past land cover are rare  and commonly used model based reconstructions
exhibit considerable differences.
Recently \citet[Spatial Statistics, 24:14--31,][]{PirzaLPG2018_24} developed a statistical
interpolation method that produces spatially complete reconstructions of past
land cover from pollen assemblage. These reconstructions incorporate a number of
auxiliary datasets raising questions regarding the method's sensitivity to
different auxiliary datasets.

Here the sensitivity of the method is examined by performing spatial
reconstructions for northern Europe during three time periods (1900 CE, 1725 CE
and 4000 BCE). The auxiliary datasets considered include the most commonly
utilized sources of past land-cover data --- e.g.\ estimates produced by a
dynamic vegetation (DVM) and anthropogenic land-cover change (ALCC) models. Five
different auxiliary datasets were considered, including different climate data
driving the DVM and different ALCC models. The resulting reconstructions were also
evaluated using cross-validation for all the time periods. For the recent time
period, 1900 CE, the different land-cover reconstructions were compared
against a present day forest map.

The validation confirms that the statistical model provides a robust spatial
interpolation tool with low sensitivity to differences in auxiliary data and
high capacity to capture information in the pollen based proxy data. Further
auxiliary data with high spatial detail improves model performance for areas
with complex topography or few observations.
\end{abstract}

\section{Introduction}
\label{sec:intro}

The importance of terrestrial land cover for the global carbon cycle and its
impact on the climate system is well recognized \citep[e.g.][]{ClausBG2001_28, BrovkCDFKLMRSS2006_26a, ArnetHZTMBFKKOo2010_3, ChrisSHB2013_40}. Many
studies have found large climatic effects associated with changes in land
cover. Forecast simulations evaluating the effects of human induced global
warming predict a considerable amplification of future climate change, especially for Arctic areas \citep{ZhangMSWKD2013_8, RichtJO2011, ChapmW2007_20,
  KoeniBGKSTWW2013_40, MilleS2012_41}. The past anthropogenic deforestation
of the temperate zone in Europe was lately demonstrated to have an impact on
regional climate similar in amplitude to present day climate change
\citep{StranKPWGTMDKBBFGKKKKKLMMNSSS2014_10}. However, studies on the effects of vegetation and
land-use changes on past climate and carbon cycle often report considerable
differences and uncertainties in their model predictions
\citep{NobleBPBBCDGHLMMRSV2012_25, PhD_Olofsson2013}.

One of the reasons for such widely diverging results could be the differences in
past land-cover descriptions used by climate modellers. Possible land-cover
descriptions range from static present-day land cover
\citep{StranBKS2011_63}, over simulated potential natural land cover from
dynamic (or static) vegetation models (DVMs)
\citep[e.g.~][]{BrovkBCGKPA2002_16, HicklVFMSCGFCCKS2012_21}, to past land-cover
scenarios combining DVM derived potential vegetation with estimates of
anthropogenic land-cover change (ALCC) \citep{StranKPWGTMDKBBFGKKKKKLMMNSSS2014_10, PongrRRC2008_22,
  NobleBPBBCDGHLMMRSV2012_25}. Differences in input climates, mechanistic and
parametrisation differences of DVMs \citep{PrentBCHHLSSS2007_Canadell,
  ScheiLH2013_198}, and significant variation between existing ALCC scenarios
\citep[e.g.~][]{KaplaKZ2009_28, PongrRRC2008_22, KleinBVD2011_20,
  GaillSMTBHKKKKLMOPRSSFNABBBBBGHKKKKLLLLLOPSSS2010_6} further contribute to the differences in past land-cover
descriptions. These differences can lead to largely diverging estimates of past
land-cover dynamics even when the most advanced models are used
\citep{StranKPWGTMDKBBFGKKKKKLMMNSSS2014_10, PitmaNCDBBCDGGHLMMRSSV2009_36}. Thus, reliable land-cover
representations are important when studying biogeophysical impacts of anthropogenic
land-cover change on climate.

The palaeoecological proxy based land-cover reconstructions recently
published by \citet{PirzaLPSTFMNKBBGKLMSG2014_20, PirzaLPG2018_24} were designed to overcome
the problems described above. And to provide a proxy based land-cover
description applicable for a range of studies on past vegetation and its
interactions with climate, soil and humans. These reconstructions use the 
pollen based land-cover composition (PbLCC) published by \citet{TrondGSMFLMNTBBBCDDDFGHKKLLLMOPPRKGSW2015_21} as
a source of information on past land-cover composition. The PbLCC are point
estimates, depicting the land-cover composition of the area surrounding each of
the studied sites. Spatial interpolation is needed to fill the gaps between observations and to produce continuous land-cover reconstructions. Conventional interpolation methods might struggle when handling noisy, spatially heterogeneous data \citep{HeuveBS1989_3,   KnegtLCSBHKSWP2010_91}, but statistical methods for spatially structured data exist \citep{GelfaDGF2010, BlangC2015}.

In \citet{PirzaLPG2018_24} a statistical model based on Gaussian Markov Random Fields \citep{LindgRL2011_73, RueH2005} was developed to provide a reliable, computationally effective and freeware based spatial interpolation technique. The resulting statistical model combines PbLCC data with auxiliary datasets; e.g.\ DVM output, ALCC scenarios, and elevation; to produce reconstructions of past land cover. 
The auxiliary data is subject to the differences and uncertainties outlined
above and the choice of auxiliary data could influence the accuracy of the statistical model. 
The major objectives of this paper are: 1) To draw attention of climate modelling community to a novel set of spatially explicit pollen-proxy based land-cover reconstructions suitable for climate modelling; 2) to present and test the robustness of the spatial interpolation model developed by \citet{PirzaLPG2018_24}; and 3) to evaluate the models capacity to recover information provided by PbLCC proxy data and to analyse its sensitivity to different auxiliary datasets.

\section{Material and Methods}
\label{sec:MaterialMethod}
\begin{figure}
  \centerline{\includegraphics[width=\textwidth]{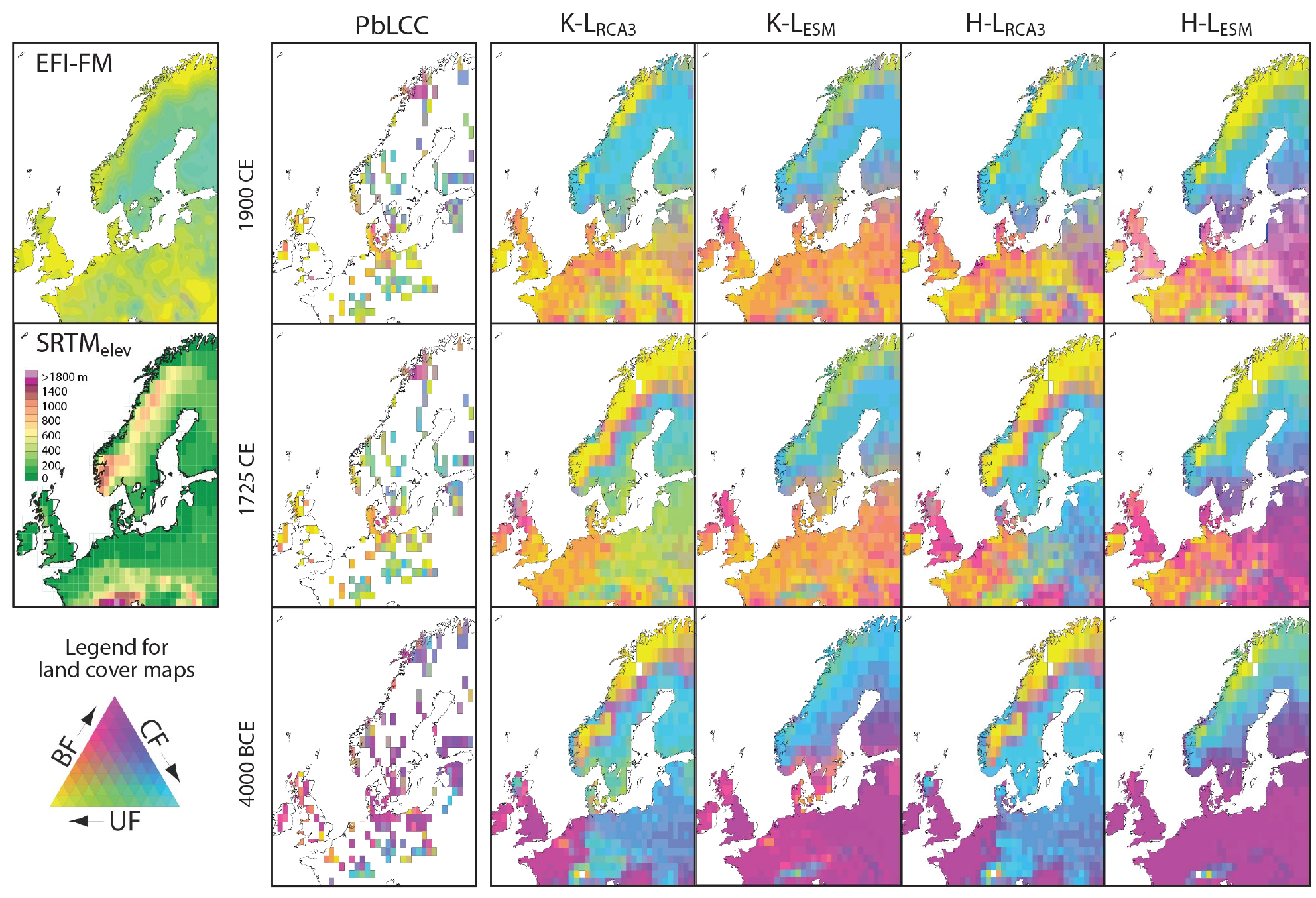}}
  \caption{Data used in the modelling. The first column shows (from top to
    bottom) the EFI-FM, SRTM$_\text{elev}$, and the colorkey for the
    land-cover compositions, coniferous forest (CF), broadleaved forest (BF) and unforested land (UF). The remaining columns give (from left to right) the
    PbLCC \citep{TrondGSMFLMNTBBBCDDDFGHKKLLLMOPPRKGSW2015_21} and
    the four model based compositions considered as potential covariates: 
    K-L$_\text{RCA3}$, K-L$_\text{ESM}$, H-L$_\text{RCA3}$, and H-L$_\text{ESM}$. 
    Here K/H indicates KK10 \citep{KaplaKZ2009_28} or HYDE \citep{KleinBVD2011_20}
     land use scenarios and L$_\text{RCA3}$/L$_\text{ESM}$ indicates the climate ---
     Rossby Centre Regional Climate Model \citep{SamueJWUGHJKNW2011_63} or Earth System Model \citep{MikolGMSVW2007_28} --- used to drive the vegetation model.
     The three rows represent (from top to bottom) the time periods 1900 CE, 1725 CE,
      and 4000 BCE.}
  \label{fig:data}
\end{figure}

The studied area covers temperate, boreal and alpine-arctic biomes of central and northern Europe ($45^\circ$N to $71^\circ$N and $10^\circ$W to $30^\circ$E). 
The PbLCC data published in \citet{TrondGSMFLMNTBBBCDDDFGHKKLLLMOPPRKGSW2015_21} consists of proportions of coniferous forest, broadleaved forest and unforested land presented as gridded ($1^\circ\times 1^\circ$) data points placed irregularly across northern-central Europe. Altogether 175 grid cells containing proxy data were available for 1900 CE, 181 for 1725 CE, and 196 for the 4000 BCE time-period (Figure~\ref{fig:data}, column 2). 

Four different model derived datasets, depicting past land cover, along with elevation were considered as potential auxiliary datasets.
In each case potential natural vegetation composition estimated by the DVM LPJ-GUESS \citep[Lund-Potsdam-Jena General Ecosystem Simulator;][]{SmithPS2001_10, SitchSPABCKLLSTV2003_9} were combined with an ALCC scenario to adjust for human land use \citep[see][for details]{PirzaLPSTFMNKBBGKLMSG2014_20}:
\begin{description}
\item[K-L$_\text{RCA3}$:] Combines the ALCC scenario KK10
  \citep{KaplaKZ2009_28} and the potential natural vegetation from LPJ-GUESS. Climate forcing for the DVM was derived from RCA3 \citep[Rossby Centre Regional
  Climate Model, ][]{SamueJWUGHJKNW2011_63} at annual time and
  $0.44^\circ\times0.44^\circ$ spatial resolution (Figure~\ref{fig:data},
  column 3),
\item[K-L$_\text{ESM}$:] Combines the ALCC scenario KK10 and the potential natural vegetation from LPJ-GUESS. Climate forcing for the DVM was derived
  from the Earth System Model \citep[ESM;][]{MikolGMSVW2007_28} at centennial
  time and  $5.6^\circ\times5.6^\circ$ spatial resolution. 
  To interpolate data into annual time and $0.5^\circ\times0.5^\circ$ spatial
  resolution climate data from 1901--1930 CE provided by the Climate Research
  Unit was used  (Figure~\ref{fig:data}, column 4),
\item[H-L$_\text{RCA3}$:] Combines the ALCC scenario from the History Database
  of the Global Environment \citep[HYDE; ][]{KleinBVD2011_20} and vegetation
  from LPJ-GUESS with RCA3 climate forcing (Figure~\ref{fig:data}, column 5),
\item[H-L$_\text{ESM}$:] Combines the ALCC scenario from HYDE and vegetation
  from LPJ-GUESS with ESM climate forcing (Figure~\ref{fig:data}, column
  6).
\end{description} 
The elevation data (denoted SRTM$_\text{elev}$) was obtained from the Shuttle Radar Topography Mission \citep{BeckeSSBBDFFIKLMNPSTVWW2009_32} (Figure~\ref{fig:data}, column 1 row 2).

Finally, a modern forest map based on data from the European Forest Institute
(EFI) is used for evaluation of the model's performance for the 1900 CE time period.
The EFI forest map (EFI-FM) is based on a combination of satellite data
and national forest-inventory statistics from 1990--2005
\citep{PaeiviLSHVKF2001_Combining, SchucBPHKF2002_Compilation} (Figure~\ref{fig:data},
column 1 row 1). All auxiliary data were up-scaled to $1^\circ \times 1^\circ$
spatial resolution, matching the pollen based reconstructions, before usage as
model input. 

\subsection{Statistical Model for Land-cover Compositions}
\label{subsec:Model}
\begin{figure}
	\centerline{\includegraphics[scale=0.6]{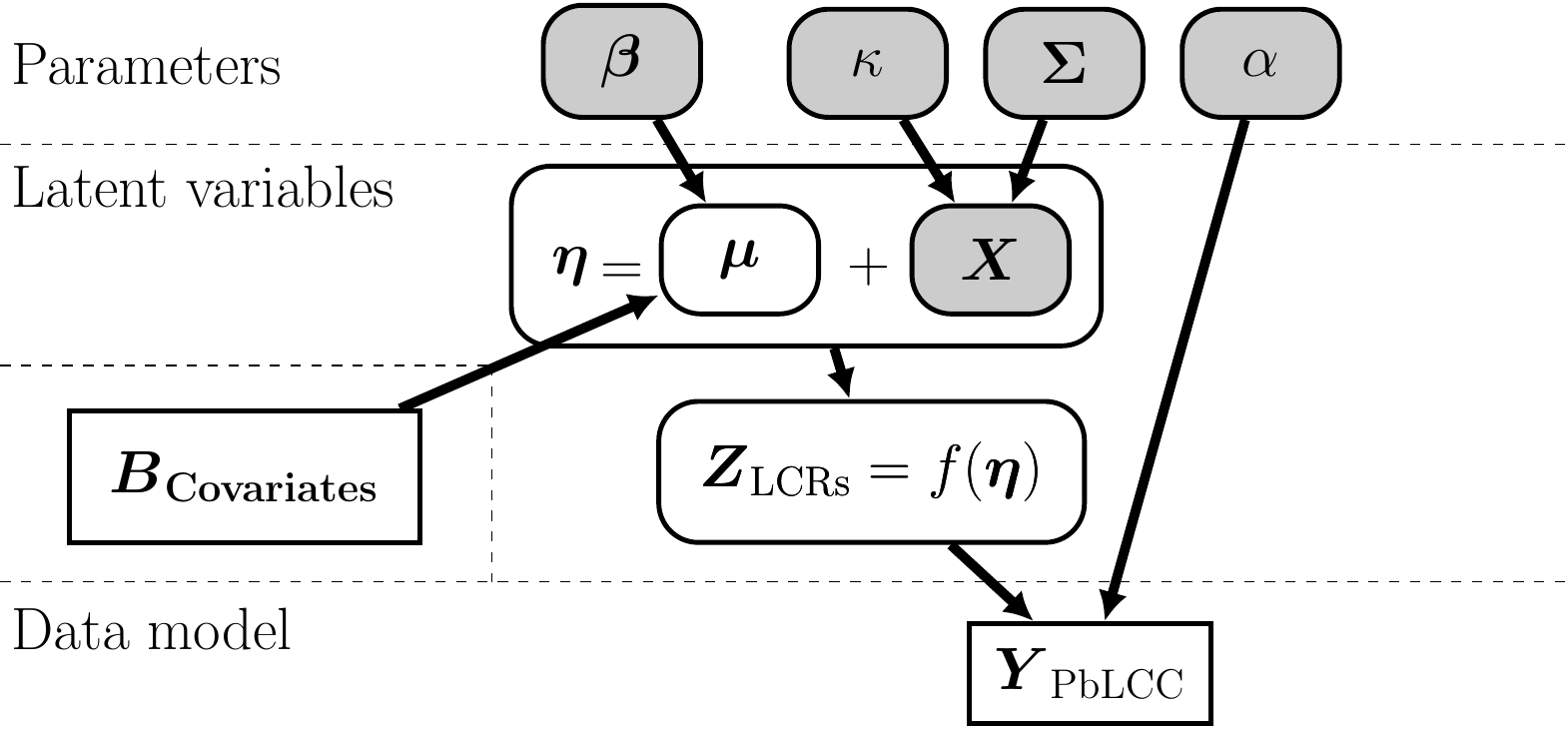}}
	\caption{Hierarchical graph describing the conditional dependencies between
		observations (white rectangle) and parameters (grey rounded rectangles) to
    be estimated. The white rounded rectangles are computed based on the
    estimations. In a generalized linear mixed model framework, $\boldsymbol{\eta}$ is
    the linear predictor --- consisting of a regression term, $\boldsymbol{\mu}$, and a
    spatial random effect, $\boldsymbol{X}$. The link function, $f(\boldsymbol{\eta})$,
    transforms between linear predictor and proportions, which are matched to
    the observed land cover proportions, $\boldsymbol{Y}_\text{PbLCC}$, using a
    Dirichlet distribution.}
	\label{fig:hierarchy}
\end{figure}

A Bayesian hierarchical model is used to interpolate the PbLCC data; here we
only provide a brief overview of the model, mathematical and technical details
can be found in \citet{PirzaLPG2018_24}. The model can be seen as a special
case of a generalized linear mixed model with a spatially correlated random
effect. An alternative interpretation of the model is as an empirical forward
model (direction of arrows in Figure~\ref{fig:hierarchy}) where parameters
affect the latent variables which in turn affect the data. Reconstructions are
obtained by inverting the model (i.e.\ computing the posterior) to obtain the
latent variables given the data.

The PbLLC derived proportions of land cover (coniferous forest, broadleaved
forest and unforested land), denoted $\boldsymbol{Y}_\text{PbLCC}$, are seen as draws
from a Dirichlet distribution \citep[Ch.~49]{KotzBJ2000} given a vector of proportions, $\boldsymbol{Z}$, and a
concentration parameter, $\alpha$ (controlling the uncertainty:
$\mathsf{V}(\boldsymbol{Y}_\text{PbLCC}) \propto 1/\alpha$). Since the proportions have
to obey certain restrictions ($0 \leq Z_k \leq 1$ and $\sum_{k=1}^3 Z_k=1$, were
$k$ indexes the land-cover types), a link function is used to transform between
the proportions and the linear predictor, $\boldsymbol{\eta}$:
\begin{align*}
	Z_k &= f(\boldsymbol{\eta}) =
	\begin{cases}
	\frac{e^{\eta_k}}{1+\sum_{i=1}^2 e^{\eta_i}} & \text{for } k = 1,2\\
	\frac{1}{1+\sum_{i=1}^2e^{\eta_i}} & \text{for } k = 3
	\end{cases}
	\\
    \eta_k &= f^{-1}(\boldsymbol{Z}) = \log\left( \frac{Z_k}{Z_3} \right) 
    \qquad \text{for } k=1,2
\end{align*}
Here $f^{-1}(\boldsymbol{Z})$ is the additive log-ratio transformation \citep{Aitch1986}, a multivariate extension of the logit transformation.

The linear predictor consists of a mean structure and a spatially dependent random effect, $\boldsymbol{\eta} = \boldsymbol{\mu} + \boldsymbol{X}$. The mean structure is modelled as a linear regression, $\boldsymbol{\mu}=\boldsymbol{B}\boldsymbol{\beta}$; i.e.\ a combination of covariates, $\boldsymbol{B}$, and regression coefficients, $\boldsymbol{\beta}$. To aid in variable selection and suppress uninformative covariates a horseshoe prior \citep{ParkC2008_103, MakalS2016_23} is used for $\boldsymbol{\beta}$. The main focus of this paper is to evaluate the model sensitivity to the choice of covariates (i.e.\ the auxiliary datasets). The PbLCC is modelled based on six different sets of covariates: 1) Intercept, 2) SRTM$_\text{elev}$, 3) K-L$_\text{ESM}$, 4) K-L$_\text{RCA3}$, 5) H-L$_\text{ESM}$, and 6) H-L$_\text{RCA3}$;  illustrated in Figure~\ref{fig:data}. A summary of the different models is given in Table~\ref{tab:Bcovariates}. 

Finally, the spatially dependent random effect is modelled using a Gaussian Markov Random Field \citep{LindgRL2011_73} with two parameters: $\kappa$, controlling the strength of the spatial dependence and $\boldsymbol{\Sigma}$, controlling the variation within and between the fields (i.e.\ the correlation among different land cover types).

\begin{table}
  \caption{Six different models and corresponding covariates. SRTM$_\text{elev}$
    is elevation \citep{BeckeSSBBDFFIKLMNPSTVWW2009_32}, K/H indicates KK10
    \citep{KaplaKZ2009_28} or HYDE \citep{KleinBVD2011_20} land use
    scenarios and L$_\text{RCA3}$/L$_\text{ESM}$ indicates vegetation model
    driven by climate from the Rossby Centre Regional Climate Model
    \citep{SamueJWUGHJKNW2011_63} or from an Earth System Model
    \citep{MikolGMSVW2007_28}.}
  \label{tab:Bcovariates}
  \begin{center}
    \begin{tabular}{l|cccccc}
      \diagbox[innerwidth=2.5cm]{Model}{Covariates} & Intercept & SRTM$_\text{elev}$ & K-L$_\text{ESM}$ & K-L$_\text{RCA3}$ & H-L$_\text{ESM}$ & H-L$_\text{RCA3}$ \\
      \hline											
      Constant  		&	x	&		&		&		&		&		\\
      Elevation			&	x	&	x	&		&		&		&		\\			
      K-L$_\text{ESM}$	&	x	&	x	&	x	&		&		&		\\
      K-L$_\text{RCA3}$	&	x	&	x	&		&	x	&		&		\\
      H-L$_\text{ESM}$	&	x	&	x	&		&		&	x	&		\\
      H-L$_\text{RCA3}$	&	x	&	x	&		&		&		&	x	\\
    \end{tabular}
  \end{center}
\end{table}

Model estimation and reconstructions are performed using Markov Chain Monte Carlo \citep{BrookGJM2011} with $100\,000$ samples and a burn-in of $10\,000$ \citep[See][for details.]{PirzaLPG2018_24}. Output from the Markov Chain Monte Carlo are then used to compute land-cover reconstructions (as posterior expectations, $\mathsf{E}(\boldsymbol{Z}|\boldsymbol{Y}_\text{PbLCC})$) and uncertainties in the form of predictive regions. The predictive regions describe the uncertainties associated with the reconstructions; including uncertainties in model parameters and linear predictor.%  To compare uncertainties of different reconstructions, we report the fraction of the unit (compositional) triangle covered by the predictive region for each model.

%\begin{figure}[htp]
%  \centerline{\includegraphics[width=\textwidth]{concept_ellipse_ratio_4}}
%  \caption{The left plot shows the $95\%$ elliptical predictive regionin $\mathbb{R}^2$. 
%  	The right ternary diagram shows the transformed $95\%$ predictive region
%    together with the corresponding fraction, $60\%$, compared to the whole
%    triangle.}
%  \label{fig:concept}
%\end{figure}

\subsection{Testing the Model Performance}
To evaluate model performance, we compared the land-cover reconstructions from different models for the 1900 CE time period with the EFI-FM by computing the average compositional distances  \citep{AitchBMP2000_32, PirzaLPG2018_24}. This measure is similar to root mean square error in $\mathbb{R}^2$ but it accounts for compositional properties (i.e.\ $0 \leq Z_k \leq 1$ and $\sum_{k=1}^3 Z_k=1$).

Since no independent observational data exists for the 1725 CE and 4000 BCE time
periods, we applied a 6-fold cross-validation scheme
\citep[][Ch. 7.10]{HastiTF2001} to all models and time periods. The
cross-validation divides the observations into 6 random groups and the
reconstruction errors for each group when using only observations from the other
5 groups are computed. To further compare predictive performance of the models
Deviance Information Criteria \citep[DIC; see Ch.~7.2 in][]{GelmaCSDVR2014} were
computed for all models and time periods. The DIC is a hierarchical
modelling generalization of the Akaike and Bayesian information criteria \citep[Ch.~7]{HastiTF2001}.

\section{Results and Discussion}
\label{sec:result}
Fossil pollen is a well-recognized information source of vegetation dynamics and generally accepted as the best observational data on past land-cover composition and environmental conditions \citep{TrondGSMFLMNTBBBCDDDFGHKKLLLMOPPRKGSW2015_21}.

Today, central and northern Europe have, at the subcontinental spatial scale, the highest density of palynologically investigated sites on Earth. However, even there the existing pollen records are irregularly placed, leaving some areas with scarce data coverage \citep{FyfeWR2015_21}. The collection of new pollen data to fill these gaps is very time consuming and cannot be performed everywhere. All this makes pollen data, in their original format, heavily underused, since the data is unsuitable for models requiring continuous land-cover representations as input. The lack of spatially explicit proxy based land cover data directly usable in climate models has been hampering the correct representation of past climate-land cover relationship.

Regrettably, the commonly used DVM derived representations of past land cover exhibit large variation in vegetation composition estimates. The model derived land-cover datasets used as auxiliary data (Table~\ref{tab:Bcovariates}) show large variation in estimated extents of coniferous and broadleaved forests, and unforested areas for all of the studied time periods (Figure \ref{fig:data}). These substantial differences illustrate large deviances between model based estimates of the past land-cover composition due to differences in applied climate forcing and/or ALCC scenarios. Differences in climate model outputs \citep{HarriBBPBHHIW2014_43, GladsRVABBKKLMOOPV2005_32} and ALCC model estimates \citep{GaillSMTBHKKKKLMOPRSSFNABBBBBGHKKKKLLLLLOPSSS2010_6} have been recognized in earlier comparison studies and syntheses. The effect of the differences in input climate forcing and ALCC scenario on DVM estimated land-cover composition presented here are especially pronounced for central and western Europe, and for elevated areas in northern Scandinavia and the Alps (Figure \ref{fig:data}). In general the KK10 ALCC scenario produces larger unforested areas, notably in western Europe, compared to the HYDE scenario. Compared to the ESM climate forcing; the RCA3 forcing results in higher proportions of coniferous forest, especially for central, northern and eastern Europe. The described differences are clearly recognizable for all the considered time periods and are generally larger between time periods than within each time period. The purpose of the statistical model presented in Section~\ref{subsec:Model} is to combine the observed PbLCC with the spatial structure in the auxiliary data to produce data driven spatially complete maps of past land-cover that can be used directly (as input) in others models.

\begin{figure}
  \centerline{\includegraphics[width=0.8\textwidth]{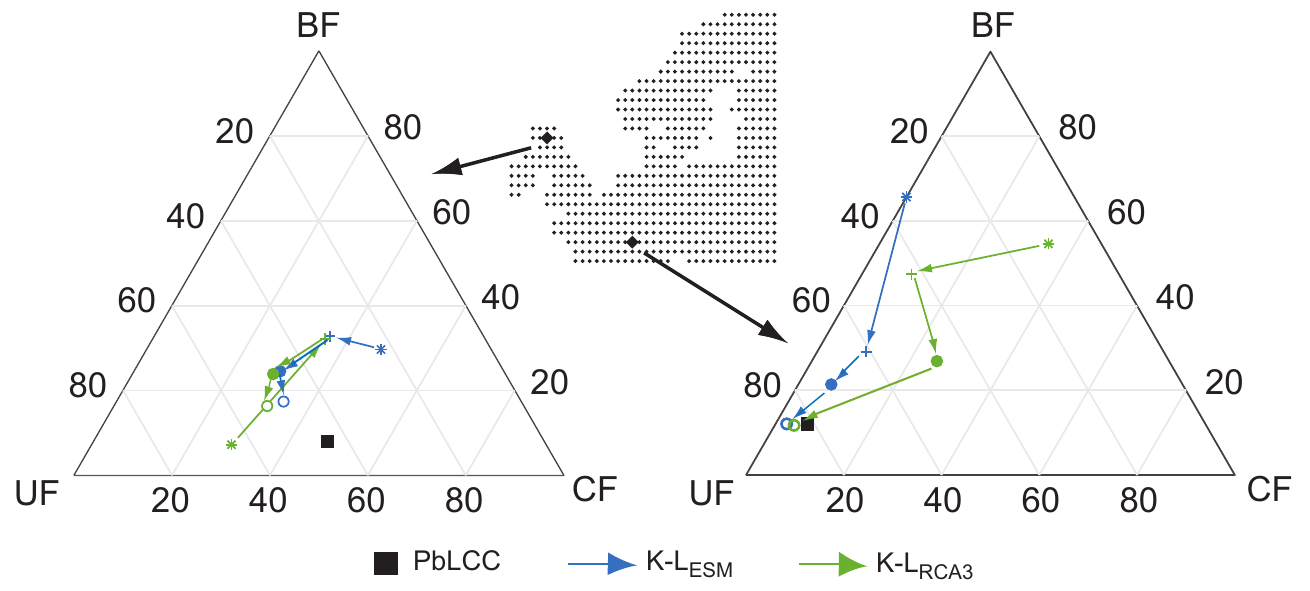}}
  \caption{Advancement of the model for two locations at 1725 CE. Starting 
		from the value of the K-L$_\text{RCA3}$ and K-L$_\text{ESM}$
		covariates ($*$), the cumulative effects of regression coefficients,
		$\boldsymbol{\beta}$, ($+$); the intercept and SRTM$_\text{elev}$ covariates
		($\bullet$); and, finally, the spatial dependency structures 
		($\circ$), are illustrated. With the final points ($\circ$) 
		corresponding to the land-cover reconstructions and $\blacksquare$ marking the observed pollen based land-cover composition.}
  \label{fig:2point_concept}
\end{figure}

\begin{figure}
  \centerline{\includegraphics[width=1.05\textwidth]{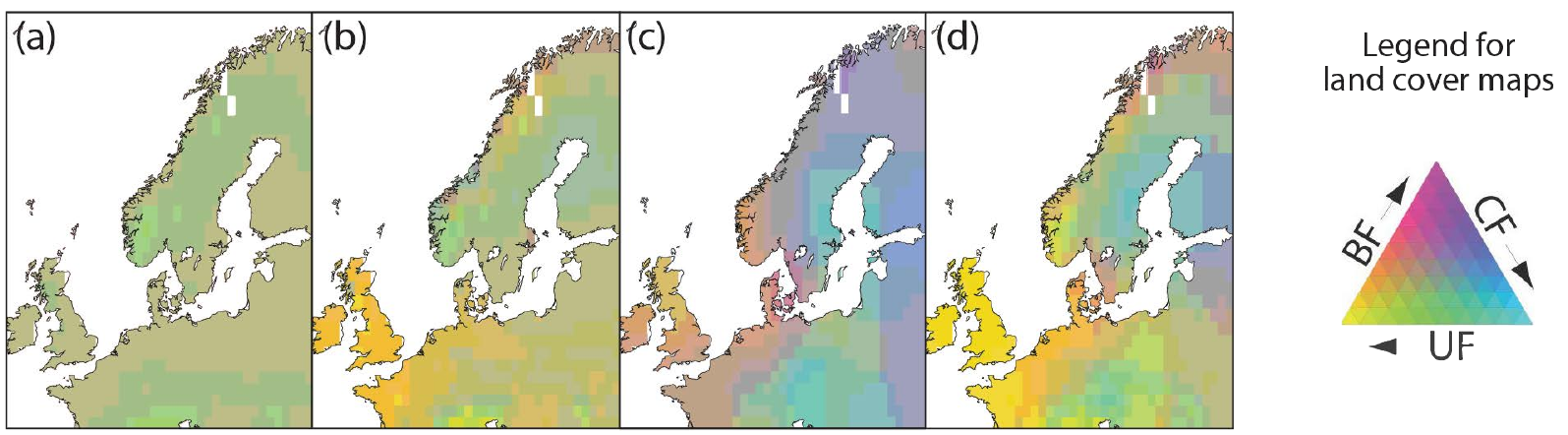}}
  \caption{Advancement of K-L$_\text{ESM}$ models for the 1725 CE time
   period: (a) shows the effect of intercept and SRTM$_\text{elev}$, (b) 
  	shows the mean structure, $\boldsymbol{\mu}$, including all the covariates, 
  	(c) shows the spatial dependency structure and finally (d) shows the
  	 resulting land-cover reconstructions obtained by
  	  adding (b) and (c).}
  \label{fig:beta_advance}
\end{figure}

To illustrate the structure of the statistical model, step by step advancement
from auxiliary data (model derived land cover) to final statistical estimates of land cover compositions for 1725 CE are given in Figures~\ref{fig:2point_concept} and
\ref{fig:beta_advance}. The large differences in K-L$_\text{RCA3}$ and
K-L$_\text{ESM}$ are reduced by scaling with the regression coefficients,
$\boldsymbol{\beta}$, capturing the empirical relationship between covariates
and PbLCC data. Thereafter, the land-cover estimates are subjected to similar
adjustments due to intercept and SRTM$_\text{elev}$, and finally similar spatial
dependent effects.

\begin{figure}
  \centerline{\includegraphics[width=\textwidth]{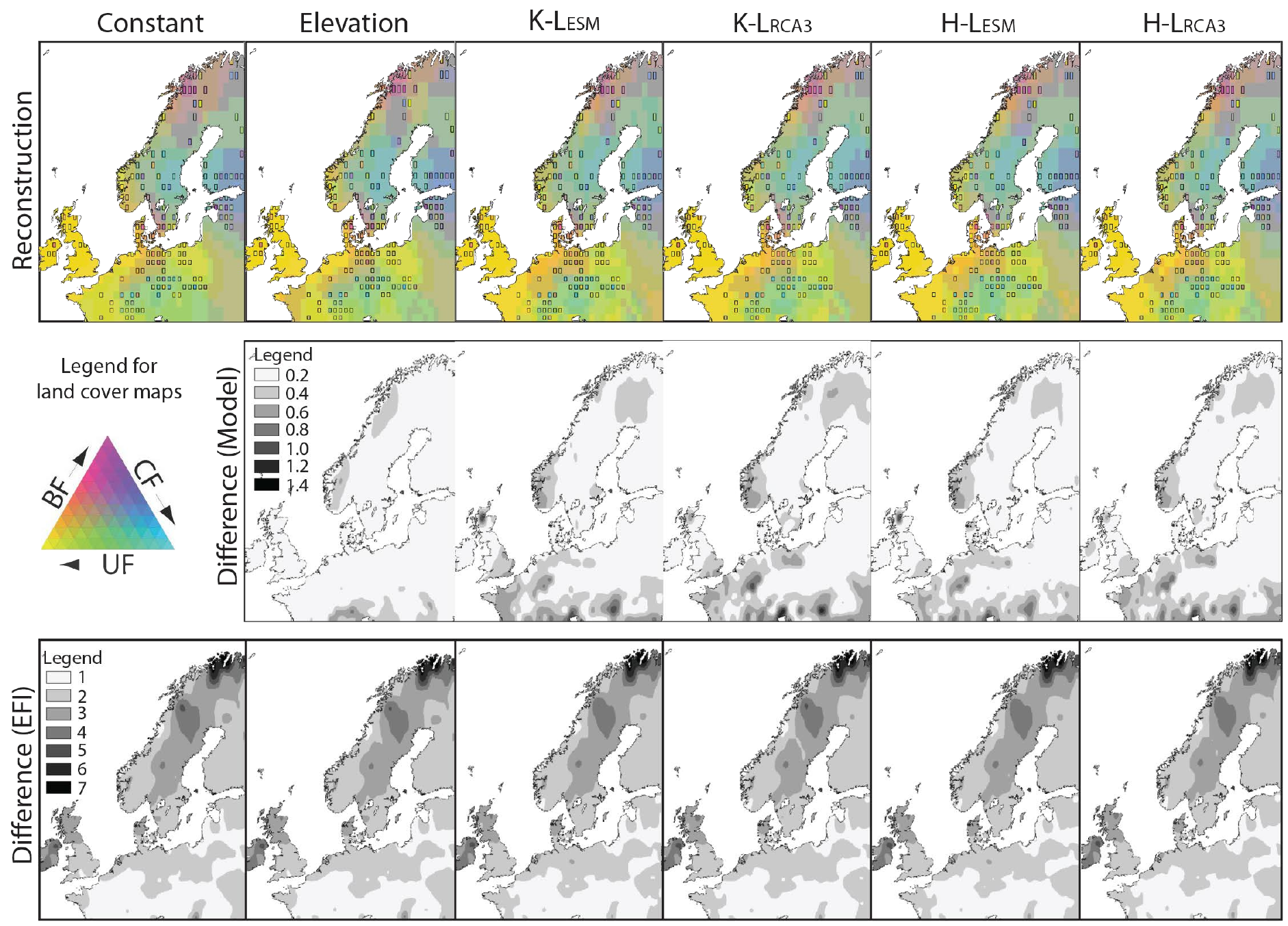}}
  \caption{Land-cover reconstructions using PbLCC for the 1900 CE time periods (top row). The reconstructions are based on six 
    different models (see Table~\ref{tab:Bcovariates}) with different auxiliary 
    datasets. Locations and compositional values of the available PbLCC data are given 
    by the black rectangles. Middle row shows the compositional distances between each model and the Constant model. Bottom row shows the compositional distances between each model and the EFI-FM.}
  \label{fig:LCR_1900}
\end{figure}

\begin{figure}
  \centerline{\includegraphics[width=\textwidth]{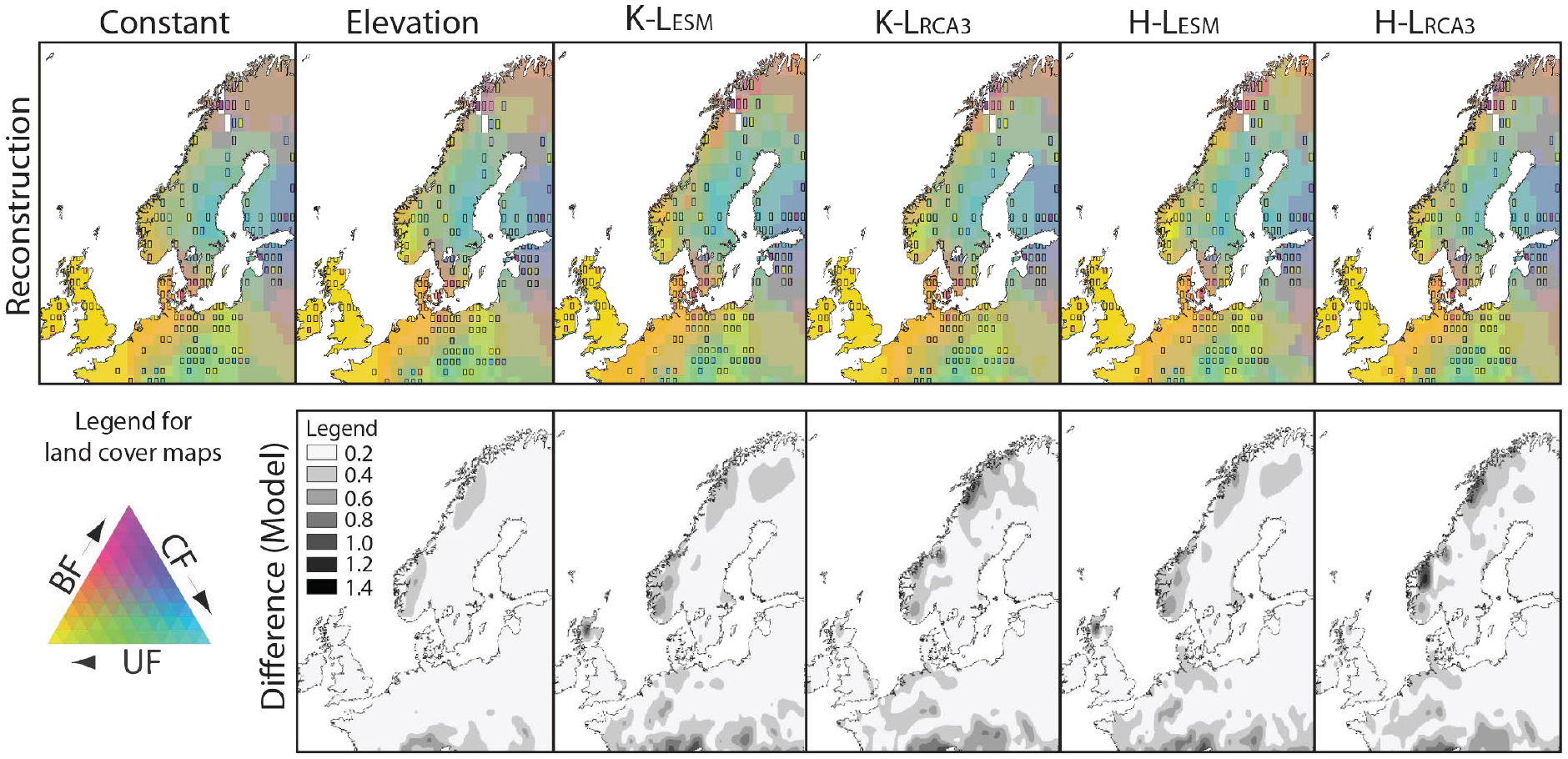}}
  \centerline{\includegraphics[width=\textwidth]{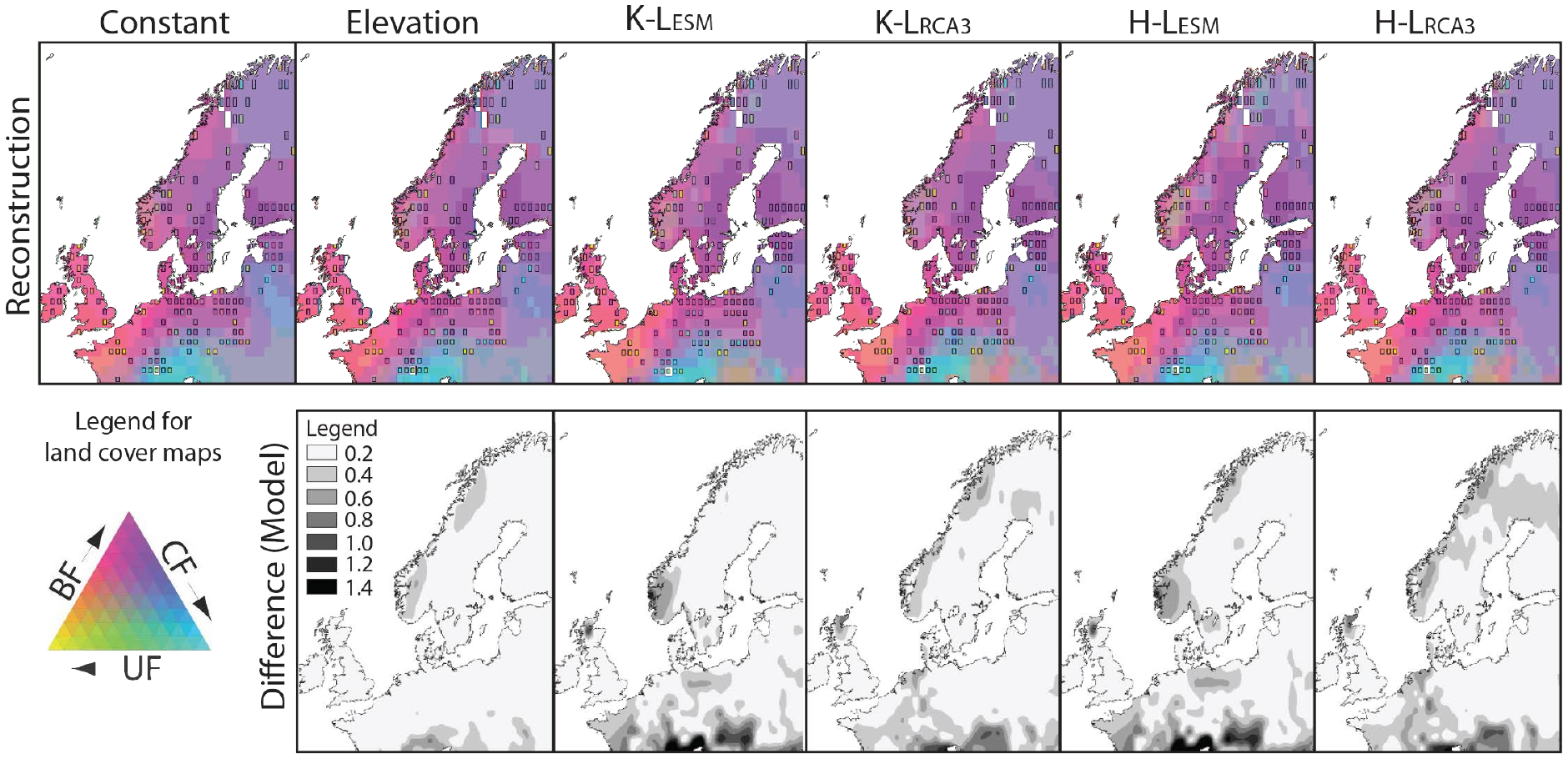}}
  \caption{Land-cover reconstructions using local estimates of PbLCC for the
    1725 CE (top) and 4000 BCE (bottom) time periods. The reconstructions are
    based on six different models (see Table~\ref{tab:Bcovariates}) with
    different auxiliary datasets. Locations and compositional values of the
    available PbLCC data are given by the black rectangles. Third and fourth row
    show the compositional distances between each model and the Constant model.}
  \label{fig:LCR_1725_4000}
\end{figure}

The impact of different auxiliary datasets was assessed by using 
the statistical model to create a set of proxy based reconstructions of past
land cover for central and northern Europe during three time periods (1900 CE,
1725 CE and 4000 BCE; see Figures~\ref{fig:LCR_1900} and \ref{fig:LCR_1725_4000}).
Each of these reconstructions were based on the irregularly distributed observed
pollen data (PbLCC), available for ca $25\%$ of the area, together with one of
the six models (Table~\ref{tab:Bcovariates}) using different combinations of the auxiliary data (Figure~\ref{fig:data}).

\begin{figure}[htp]
  \noindent\makebox[\textwidth]{
    \includegraphics[width=\textwidth]{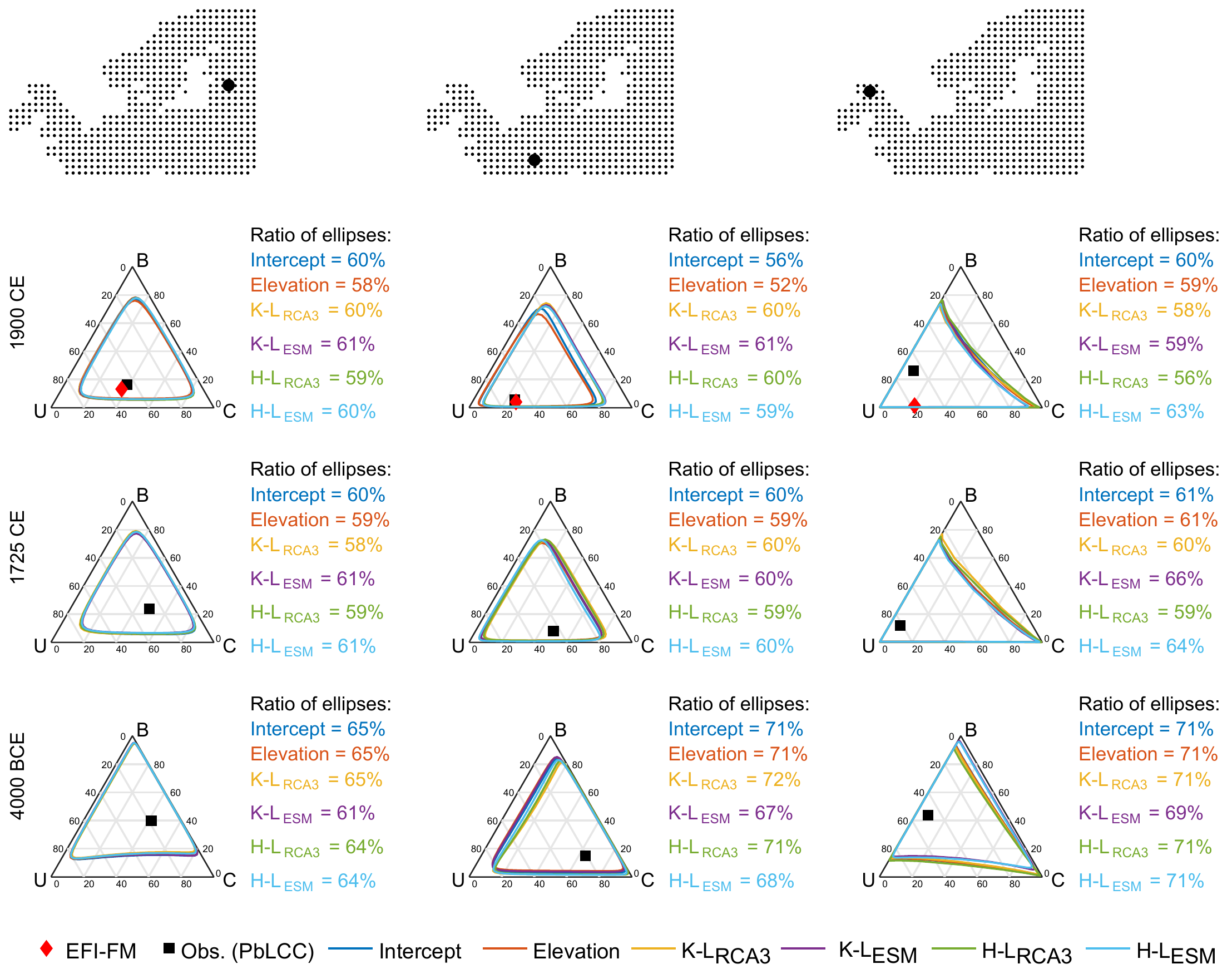}}
  \caption{The prediction regions and fraction of the ternary triangle covered
    by these regions are presented for three locations, the six models, and the 
    1900 CE, 1725 CE and 4000 BCE time periods.}
  \label{fig:PR}
\end{figure}

The resulting land-cover reconstructions exhibit considerably higher similarity
with the PbLCC data than any of the auxiliary land-cover datasets for all tested models
and time periods (Figures~\ref{fig:LCR_1900} and \ref{fig:LCR_1725_4000}).
At first the similarity among the reconstructions might seem contradictory, but
recall that the model allows for, and estimates, different weighting (the
regression coefficients, $\boldsymbol{\beta}$:s) for each of the auxiliary
datasets. Thus, the resulting reconstructions do not rely on the absolute values
in the auxiliary datasets, only their spatial patterns. As a result, model
performance for elevated areas and for the areas with low observational data
coverage (e.g.\ eastern and south-eastern Europe) is improved by including
covariates that exhibit distinct spatial structures for the given areas
(Figures~\ref{fig:LCR_1900} and \ref{fig:LCR_1725_4000}).
Neither the DIC results nor the 6-fold cross-validation results show any
advantage among the six tested models for the different time periods
(Table~\ref{tab:DIC_ACD}). Analogous to the reconstructions, the predictive
regions are very similar in both size and shape irrespective of the auxiliary
dataset used, indicating similar reconstruction uncertainties across all models
(Figure~\ref{fig:PR}). Implying there is no clear preference among the models,
i.e.\ that the results are robust to the choice of auxiliary dataset.

\begin{table}[htp]
 \begin{center}
 	\begin{tabular}{c|ccc||ccc}
 	& \multicolumn{3}{c||}{DIC} & \multicolumn{3}{c}{ACD} \\
 	& 1900CE& 1725CE& 4000BCE & 1900CE& 1725CE& 4000BCE\\
 	\hline
       Constant			&	-559	&	-655	&	-593	&	1.00	&	1.12	&	1.20	\\
	  Elevation			&	\textbf{-568}	&	-664	&	-589	&	0.99	&	\textbf{1.11}	&	1.21	\\
	  K-L$_\text{ESM}$	&	-547	&	-649	&	\textbf{-609}	&	1.00	&	1.12	&	1.18	\\
	  K-L$_\text{RCA3}$	&	-549	&	-661	&	-589	&	0.99	&	1.13	&	1.19	\\
	  H-L$_\text{ESM}$	&	-549	&	-655	&	-608	&	0.99	&	1.11	&	\textbf{1.17}	\\
	  H-L$_\text{RCA3}$	&	-557	&	\textbf{-669}	&	-595	&	\textbf{0.99}	&	1.12	&	1.18	\\
 	\end{tabular}
 \end{center}
 \caption{Deviance information criteria (DIC) and Average compositional distances (ACD) from 6-fold cross-validations for each of the six models and three time periods. Best value for each time period marked in \textbf{bold-font}.}
 \label{tab:DIC_ACD}
\end{table}

Although a temporal misalignment exists between the PbLCC data for the 1900 CE
time period (based on pollen data from 1850 to the present) and the EFI-FM
(inventory and satellite data from 1990-2005); EFI-FM provides the best complete
and consistent land cover map of Europe for present time, making it a reasonable
choice for a comparison. The main differences between the EFI-FM and the PbLCC
data for the 1900 CE time period are: 1) lower abundance of broadleaved forests
for most of Europe, 2) higher abundance of coniferous forest in Sweden and
Finland, and 3) higher abundance of unforested land in North Norway in the
EFI-FM data than in the PbLCC data \citep{PirzaLPG2018_24}.
The average compositional distances computed between the land-cover
reconstructions and the EFI-FM for 1900 CE show practically identical ($1.47$ to
$1.48$) distances between all six reconstructions and the EFI-FM, and small
differences among the six presented models (Table~\ref{tab:ACD}).

These results clearly show that the developed statistical interpolation model is
robust to the choice of covariates. The model is suitable for reconstructing
spatially continuous maps of past land cover from scattered and irregularly
spaced pollen based proxy data.

 \begin{table}
 \begin{center}
 \begin{tabular}{l|ccccccc}
 &  EFI-FM & Elevation & K-L$_\text{ESM}$ & K-L$_\text{RCA3}$ & H-L$_\text{ESM}$ & H-L$_\text{RCA3}$ \\
 \hline
 &\multicolumn{6}{c}{1900 CE}\\
 \hline
 		Constant			&	1.48	&	0.08&	0.18&	0.20&	0.17&	0.19\\
		Elevation			&	1.49	&		&	0.19&	0.21&	0.18&	0.20\\
		K-L$_\text{ESM}$	&	1.48	&		&		&	0.09&	0.07&	0.09\\
		K-L$_\text{RCA3}$	&	1.48	&		&		&		&	0.11&	0.06\\
		H-L$_\text{ESM}$	&	1.48	&		&		&		&		&	0.08\\
		H-L$_\text{RCA3}$	&	1.48	&		&		&		&		&		\\
 \hline													
 &\multicolumn{6}{c}{1725 CE}\\					
 \hline													
 		Constant			&	0.10&	0.16&	0.16&	0.17&	0.17	\\
		Elevation			&		&	0.14&	0.11&	0.14&	0.13	\\
		K-L$_\text{ESM}$	&		&		&	0.14&	0.06&	0.16	\\
		K-L$_\text{RCA3}$	&		&		&		&	0.15&	0.07	\\
		H-L$_\text{ESM}$	&		&		&		&		&	0.15	\\
 \hline													
 &\multicolumn{6}{c}{4000 BCE}\\					
 \hline													
 		Constant			&	0.11&	0.21&	0.17&	0.22&	0.19	\\
		Elevation			&		&	0.19&	0.12&	0.20&	0.15	\\
		K-L$_\text{ESM}$	&		&		&	0.19&	0.07&	0.21	\\
		K-L$_\text{RCA3}$	&		&		&		&	0.18&	0.07	\\
		H-L$_\text{ESM}$	&		&		&		&		&	0.20	\\
 \end{tabular}
 \end{center}
 \caption{The average compositional distances among the six models fitted to the
   data for each of the three time periods.}
 \label{tab:ACD}
 \end{table}

\section{Conclusions}
\label{sec:conclusion}
The statistical model and Bayesian interpolation method presented here has been specially designed for handling irregularly spaced palaeo-proxy records like pollen data and, dependent on proxy data availability, is globally applicable. 
%The ability of the model to create pollen based land cover reconstructions at sub-continental scale was illustrated using an example application on two Holocene time periods frequently assessed by climate researchers: the Little Ice Age (1725 CE) and the Holocene Thermal Maximum (4000 BCE); as well as a third time period covering the recent past (1900 CE) which was used for validation. 
The model produces land-cover maps by combining irregularly distributed pollen based estimates of land cover with auxiliary data and a statistical model for spatial structure. The resulting maps capture important features in the pollen proxy data and are reasonably insensitive to the use of different auxiliary datasets.

Auxiliary datasets considered were complied from commonly utilized sources of past land-cover data (outputs from a dynamic vegetation model using different climatic drivers and anthropogenic land-cover changes scenarios). These datasets exhibit considerable differences in their recreation of the past land cover. Emphasizing the need for the independent, proxy based past land-cover maps created in this paper.  

Evaluation of the model's sensitivity indicates that the proposed statistical model is robust to the choice of auxiliary data and only considers features in the auxiliary data that are consistent with the proxy data. However, auxiliary data with detailed spatial information considerably improves the interpolation results for areas with low proxy data coverage, with no reduction in overall performance.
%It was shown that the model can provide reliable results using a variety of auxiliary land cover data as long as the data captures important spatial patterns; absent good auxiliary data reasonable results can be obtained using only elevation.

This modelling approach has demonstrated a clear capacity to produce empirically based land-cover reconstructions for climate modelling purposes. Such reconstructions are necessary to evaluate anthropogenic land-cover change scenarios currently used in climate modelling and to study past interactions between land cover and climate with greater reliability. The model will also be very useful for producing reconstructions of past land cover from the global pollen proxy data currently being produced by the PAGES (Past Global changES) LandCover6k initiative\footnote{\url{www.pastglobalchanges.org/ini/wg/landcover6k/intro}}.

\section{Data availability}
The database containing the reconstructions of coniferous forest, broadleaved forest and unforested land, three fractions of land cover, for the three time-periods presented in this paper, along with reconstructions for 1425 CE and 1000 BCE using only the K-L$_\text{ESM}$ are available for download from \url{https://github.com/BehnazP/SpatioCompo}. In addition the source code is available in the same repository under the open source GNU General Public License.  

\section*{Acronyms}
\begin{Description}[margin= LPJ-GUESS,nosep]
\item[DVM] Dynamical vegetation model.
\item[ALCC] Anthropogenic land-cover change.
\item[PbLCC] Pollen based land-cover composition.
\item[LPJ-GUESS] The Lund-Potsdam-Jena General Ecosystem Simulator, a DVM.
\item[EFI-FM] European Forest Institute forest map.
\end{Description}

\section*{Notation}
\begin{Description}[margin= $\boldsymbol{Y}_\text{PbLCC}$,nosep]
\item[$\boldsymbol{Y}_\text{PbLCC}$] Observations, as proportions.
\item[$f$] Link function, transforming between proportions and linear predictor.
\item[$\boldsymbol{\eta}$] Linear predictor, $\boldsymbol{\eta}=\boldsymbol{\mu}+\boldsymbol{X}$.
\item[$\boldsymbol{\mu}$] Mean structure; modelled as $\boldsymbol{\mu} = \boldsymbol{B}\boldsymbol{\beta}$ using covariates, $\boldsymbol{B}$, and regression coefficients, $\boldsymbol{\beta}$.
\item[$\boldsymbol{X}$] Spatially dependent random effect.
\item[$\alpha$] Concentrated parameter of the Dirichlet distribution (i.e.\ observational uncertainty)
\item[$\boldsymbol{\Sigma}$] Covariance matrix that determines the variation between and within fields
\item[$\kappa$] Scale parameter controlling the range of spatial dependency
\end{Description}

\section*{Acknowledgements}
The research presented in this paper is a contribution to the two Swedish strategic research areas Biodiversity and Ecosystems in a Changing Climate (BECC), and ModElling the Regional and Global Earth system (MERGE). The paper is also a contribution to PAGES LandCover6k. Lindstr\"{o}m has been funded by Swedish Research Council (SRC, Vetenskapsr{\aa}det) grant no 2012-5983. Poska has been funded by SRC grant no 2016-03617 and the Estonian Ministry of Education grant IUT1-8. The authors would like to acknowledge Marie-Jos\'{e} Gaillard for her efforts in providing the pollen based land-cover proxy data and thank her for valuable comments on this manuscript.

\bibliographystyle{abbrvnat}
\bibliography{Jabrv,IEEEabrv,allRefs}

\begin{thebibliography}{49}
\providecommand{\natexlab}[1]{#1}
\providecommand{\url}[1]{\texttt{#1}}
\expandafter\ifx\csname urlstyle\endcsname\relax
  \providecommand{\doi}[1]{doi: #1}\else
  \providecommand{\doi}{doi: \begingroup \urlstyle{rm}\Url}\fi

\bibitem[Aitchison(1986)]{Aitch1986}
J.~Aitchison.
\newblock \emph{The statistical analysis of compositional data}.
\newblock Chapman \& Hall, Ltd., 1986.

\bibitem[Aitchison et~al.(2000)Aitchison, Barcel{\'o}-Vidal,
  Mart{\'\i}n-Fern{\'a}ndez, and Pawlowsky-Glahn]{AitchBMP2000_32}
J.~Aitchison, C.~Barcel{\'o}-Vidal, J.~Mart{\'\i}n-Fern{\'a}ndez, and
  V.~Pawlowsky-Glahn.
\newblock Logratio analysis and compositional distance.
\newblock \emph{Math. Geol.}, 32\penalty0 (3):\penalty0 271--275, 2000.

\bibitem[Arneth et~al.(2010)Arneth, Harrison, Zaehle, Tsigaridis, Menon,
  Bartlein, Feichter, Korhola, Kulmala, O'donnell,
  et~al.]{ArnetHZTMBFKKOo2010_3}
A.~Arneth, S.~P. Harrison, S.~Zaehle, K.~Tsigaridis, S.~Menon, P.~J. Bartlein,
  J.~Feichter, A.~Korhola, M.~Kulmala, D.~O'donnell, et~al.
\newblock Terrestrial biogeochemical feedbacks in the climate system.
\newblock \emph{Nature Geosci.}, 3\penalty0 (8):\penalty0 525--532, 2010.
\newblock \doi{10.1038/ngeo905}.

\bibitem[Becker et~al.(2009)Becker, Sandwell, Smith, Braud, Binder, Depner,
  Fabre, Factor, Ingalls, Kim, Ladner, Marks, Nelson, Pharaoh, Sharman,
  Trimmer, VonRosenburg, Wallace, and
  Weatherall]{BeckeSSBBDFFIKLMNPSTVWW2009_32}
J.~J. Becker, D.~T. Sandwell, W.~H.~F. Smith, J.~Braud, B.~Binder, J.~Depner,
  D.~Fabre, J.~Factor, S.~Ingalls, S.~H. Kim, R.~Ladner, K.~Marks, S.~Nelson,
  A.~Pharaoh, G.~Sharman, R.~Trimmer, J.~VonRosenburg, G.~Wallace, and
  P.~Weatherall.
\newblock Global bathymetry and elevation data at 30 arc seconds resolution:
  {SRTM30\_PLUS}.
\newblock \emph{Marine Geol.}, 32\penalty0 (4):\penalty0 355--371, 2009.

\bibitem[Blangiardo and Cameletti(2015)]{BlangC2015}
M.~Blangiardo and M.~Cameletti.
\newblock \emph{Spatial and Spatio-temporal Bayesian Models with R-INLA}.
\newblock Wiley, 2015.

\bibitem[Brooks et~al.(2011)Brooks, Gelman, Jones, and Meng]{BrookGJM2011}
S.~Brooks, A.~Gelman, G.~L. Jones, and X.-L. Meng.
\newblock \emph{Handbook of {Markov} {Chain} {Monte} {Carlo}}.
\newblock CRC Press, 2011.

\bibitem[Brovkin et~al.(2002)Brovkin, Bendtsen, Claussen, Ganopolski, Kubatzki,
  Petoukhov, and Andreev]{BrovkBCGKPA2002_16}
V.~Brovkin, J.~Bendtsen, M.~Claussen, A.~Ganopolski, C.~Kubatzki, V.~Petoukhov,
  and A.~Andreev.
\newblock Carbon cycle, vegetation, and climate dynamics in the holocene:
  Experiments with the {CLIMBER-2} model.
\newblock \emph{Glob. Biogeochem. Cycles}, 16\penalty0 (4):\penalty0 1139,
  2002.

\bibitem[Brovkin et~al.(2006)Brovkin, Claussen, Driesschaert, Fichefet,
  Kicklighter, Loutre, Matthews, Ramankutty, Schaeffer, and
  Sokolov]{BrovkCDFKLMRSS2006_26a}
V.~Brovkin, M.~Claussen, E.~Driesschaert, T.~Fichefet, D.~Kicklighter,
  M.~Loutre, H.~Matthews, N.~Ramankutty, M.~Schaeffer, and A.~Sokolov.
\newblock Biogeophysical effects of historical land cover changes simulated by
  six {Earth} system models of intermediate complexity.
\newblock \emph{Clim. Dyn.}, 26\penalty0 (6):\penalty0 587--600, 2006.
\newblock \doi{10.1007/s00382-005-0092-6}.

\bibitem[Chapman and Walsh(2007)]{ChapmW2007_20}
W.~L. Chapman and J.~E. Walsh.
\newblock Simulations of {A}rctic temperature and pressure by global coupled
  models.
\newblock \emph{J. Clim.}, 20\penalty0 (4):\penalty0 609--632, 2007.
\newblock \doi{10.1175/JCLI4026.1}.

\bibitem[Christidis et~al.(2013)Christidis, Stott, Hegerl, and
  Betts]{ChrisSHB2013_40}
N.~Christidis, P.~A. Stott, G.~C. Hegerl, and R.~A. Betts.
\newblock The role of land use change in the recent warming of daily extreme
  temperatures.
\newblock \emph{Geophys. Res. Lett.}, 40\penalty0 (3):\penalty0 589--594, 2013.
\newblock \doi{10.1002/grl.50159}.

\bibitem[Claussen et~al.(2001)Claussen, Brovkin, and
  Ganopolski]{ClausBG2001_28}
M.~Claussen, V.~Brovkin, and A.~Ganopolski.
\newblock Biogeophysical versus biogeochemical feedbacks of large-scale land
  cover change.
\newblock \emph{Geophys. Res. Lett.}, 28\penalty0 (6):\penalty0 1011--1014,
  2001.

\bibitem[de~Knegt et~al.(2010)de~Knegt, van Langevelde, Coughenour, Skidmore,
  de~Boer, Heitk{\"o}nig, Knox, Slotow, van~der Waal, and
  Prins]{KnegtLCSBHKSWP2010_91}
H.~J. de~Knegt, F.~van Langevelde, M.~B. Coughenour, A.~K. Skidmore, W.~F.
  de~Boer, I.~M.~A. Heitk{\"o}nig, N.~M. Knox, R.~Slotow, C.~van~der Waal, and
  H.~H.~T. Prins.
\newblock Spatial autocorrelation and the scaling of species--environment
  relationships.
\newblock \emph{Ecology}, 91\penalty0 (8):\penalty0 2455--2465, 2010.
\newblock \doi{10.1890/09-1359.1}.

\bibitem[de~Noblet-Ducoudr{\'e} et~al.(2012)de~Noblet-Ducoudr{\'e}, Boisier,
  Pitman, Bonan, Brovkin, Cruz, Delire, Gayler, van~den Hurk, Lawrence, van~der
  Molen, M{\"u}ller, Reick, Strengers, , and
  Voldoire]{NobleBPBBCDGHLMMRSV2012_25}
N.~de~Noblet-Ducoudr{\'e}, J.-P. Boisier, A.~Pitman, G.~Bonan, V.~Brovkin,
  F.~Cruz, C.~Delire, V.~Gayler, B.~van~den Hurk, P.~Lawrence, M.~K. van~der
  Molen, C.~M{\"u}ller, C.~H. Reick, B.~J. Strengers, , and A.~Voldoire.
\newblock Determining robust impacts of land-use-induced land cover changes on
  surface climate over {North America and Eurasia}: results from the first set
  of {LUCID} experiments.
\newblock \emph{J. Clim.}, 25\penalty0 (9):\penalty0 3261--3281, 2012.
\newblock \doi{10.1175/JCLI-D-11-00338.1}.

\bibitem[Fyfe et~al.(2015)Fyfe, Woodbridge, and Roberts]{FyfeWR2015_21}
R.~M. Fyfe, J.~Woodbridge, and N.~Roberts.
\newblock From forest to farmland: pollen-inferred land cover change across
  {Europe} using the pseudobiomization approach.
\newblock \emph{Glob. Change Biol.}, 21\penalty0 (3):\penalty0 1197--1212,
  2015.
\newblock \doi{10.1111/gcb.12776}.

\bibitem[Gaillard et~al.(2010)Gaillard, Sugita, Mazier, Trondman, Brostrom,
  Hickler, Kaplan, Kjellstr{\"o}m, Kokfelt, Kune{\v{s}}, , Lemmen, Miller,
  Olofsson, Poska, Rundgren, Smith, Strandberg, Fyfe, Nielsen, Alenius,
  Balakauskas, Barnekov, Birks, Bjune, Bj{\"o}rkman, Giesecke, Hjelle, Kalnina,
  Kangur, van~der Knaap, Koff, Lager{\aa}s, Lata{\l}owa, Leydet, Lechterbeck,
  Lindbladh, Odgaard, Peglar, Segerstr{\"o}m, von Stedingk, and
  Sepp{\"a}]{GaillSMTBHKKKKLMOPRSSFNABBBBBGHKKKKLLLLLOPSSS2010_6}
M.-J. Gaillard, S.~Sugita, F.~Mazier, A.-K. Trondman, A.~Brostrom, T.~Hickler,
  J.~O. Kaplan, E.~Kjellstr{\"o}m, U.~Kokfelt, P.~Kune{\v{s}}, , C.~Lemmen,
  P.~Miller, J.~Olofsson, A.~Poska, M.~Rundgren, B.~Smith, G.~Strandberg,
  R.~Fyfe, A.~Nielsen, T.~Alenius, L.~Balakauskas, L.~Barnekov, H.~Birks,
  A.~Bjune, L.~Bj{\"o}rkman, T.~Giesecke, K.~Hjelle, L.~Kalnina, M.~Kangur,
  W.~van~der Knaap, T.~Koff, P.~Lager{\aa}s, M.~Lata{\l}owa, M.~Leydet,
  J.~Lechterbeck, M.~Lindbladh, B.~Odgaard, S.~Peglar, U.~Segerstr{\"o}m,
  H.~von Stedingk, and H.~Sepp{\"a}.
\newblock Holocene land-cover reconstructions for studies on land cover-climate
  feedbacks.
\newblock \emph{Clim. Past}, 6:\penalty0 483--499, 2010.

\bibitem[Gelfand et~al.(2010)Gelfand, Diggle, Guttorp, and
  Fuentes]{GelfaDGF2010}
A.~Gelfand, P.~J. Diggle, P.~Guttorp, and M.~Fuentes.
\newblock \emph{Handbook of spatial statistics}.
\newblock CRC Press, 2010.

\bibitem[Gelman et~al.(2014)Gelman, Carlin, Stern, Dunson, Vehtari, and
  Rubin]{GelmaCSDVR2014}
A.~Gelman, J.~B. Carlin, H.~S. Stern, D.~Dunson, A.~Vehtari, and D.~B. Rubin.
\newblock \emph{Bayesian Data Analysis}.
\newblock Chapman \& Hall/CRC, third edition, 2014.

\bibitem[Gladstone et~al.(2005)Gladstone, Ross, Valdes, Abe-Ouchi, Braconnot,
  Brewer, Kageyama, Kitoh, Legrande, Marti, R., B., R., and
  G.]{GladsRVABBKKLMOOPV2005_32}
R.~M. Gladstone, I.~Ross, P.~J. Valdes, A.~Abe-Ouchi, P.~Braconnot, S.~Brewer,
  M.~Kageyama, A.~Kitoh, A.~Legrande, O.~Marti, O.~R., O.-B. B., P.~W. R., and
  V.~G.
\newblock Mid-{H}olocene {NAO}: A {PMIP2} model intercomparison.
\newblock \emph{Geophys. Res. Lett.}, 32\penalty0 (16):\penalty0 L16707, 2005.
\newblock \doi{10.1029/2005GL023596}.

\bibitem[Goldewijk et~al.(2011)Goldewijk, Beusen, Van~Drecht, and
  De~Vos]{KleinBVD2011_20}
K.~K. Goldewijk, A.~Beusen, G.~Van~Drecht, and M.~De~Vos.
\newblock The {HYDE 3.1} spatially explicit database of human-induced global
  land-use change over the past 12,000 years.
\newblock \emph{Glob. Ecol. Biogeogr.}, 20\penalty0 (1):\penalty0 73--86, 2011.

\bibitem[Harrison et~al.(2014)Harrison, Bartlein, Brewer, Prentice, Boyd,
  Hessler, Holmgren, Izumi, and Willis]{HarriBBPBHHIW2014_43}
S.~P. Harrison, P.~J. Bartlein, S.~Brewer, I.~C. Prentice, M.~Boyd, I.~Hessler,
  K.~Holmgren, K.~Izumi, and K.~Willis.
\newblock Climate model benchmarking with glacial and mid-{H}olocene climates.
\newblock \emph{Clim. Dyn.}, 43\penalty0 (3--4):\penalty0 671--688, 2014.
\newblock \doi{10.1007/s00382-013-1922-6}.

\bibitem[Hastie et~al.(2001)Hastie, Tibshirani, and Friedman]{HastiTF2001}
T.~Hastie, R.~Tibshirani, and J.~Friedman.
\newblock \emph{The Elements of Statistical Learning}.
\newblock Springer Series in Statistics. Springer New York Inc., New York, NY,
  USA, 2001.

\bibitem[Heuvelink et~al.(1989)Heuvelink, Burrough, and Stein]{HeuveBS1989_3}
G.~B.~M. Heuvelink, P.~A. Burrough, and A.~Stein.
\newblock Propagation of errors in spatial modelling with {GIS}.
\newblock \emph{Int. J. Geogr. Inf. Syst.}, 3\penalty0 (4):\penalty0 303--322,
  1989.
\newblock \doi{10.1080/02693798908941518}.

\bibitem[Hickler et~al.(2012)Hickler, Vohland, Feehan, Miller, Smith, Costa,
  Giesecke, Fronzek, Carter, Cramer, K{\"u}hn, and
  Sykes]{HicklVFMSCGFCCKS2012_21}
T.~Hickler, K.~Vohland, J.~Feehan, P.~A. Miller, B.~Smith, L.~Costa,
  T.~Giesecke, S.~Fronzek, T.~R. Carter, W.~Cramer, I.~K{\"u}hn, and M.~T.
  Sykes.
\newblock Projecting the future distribution of {European} potential natural
  vegetation zones with a generalized, tree species-based dynamic vegetation
  model.
\newblock \emph{Glob. Ecol. Biogeogr.}, 21\penalty0 (1):\penalty0 50--63, 2012.
\newblock \doi{10.1111/j.1466-8238.2010.00613.x}.

\bibitem[Kaplan et~al.(2009)Kaplan, Krumhardt, and Zimmermann]{KaplaKZ2009_28}
J.~O. Kaplan, K.~M. Krumhardt, and N.~Zimmermann.
\newblock The prehistoric and preindustrial deforestation of {Europe}.
\newblock \emph{Quat. Sci. Rev.}, 28\penalty0 (27):\penalty0 3016--3034, 2009.

\bibitem[Koenigk et~al.(2013)Koenigk, Brodeau, Graversen, Karlsson, Svensson,
  Tjernstr{\"o}m, Will{\'e}n, and Wyser]{KoeniBGKSTWW2013_40}
T.~Koenigk, L.~Brodeau, R.~G. Graversen, J.~Karlsson, G.~Svensson,
  M.~Tjernstr{\"o}m, U.~Will{\'e}n, and K.~Wyser.
\newblock Arctic climate change in 21st century {CMIP5} simulations with
  {EC}-{Earth}.
\newblock \emph{Clim. Dyn.}, 40\penalty0 (11-12):\penalty0 2719--2743, 2013.
\newblock \doi{10.1007/s00382-012-1505-y}.

\bibitem[Kotz et~al.(2000)Kotz, Balakrishnan, and Johnson]{KotzBJ2000}
S.~Kotz, N.~Balakrishnan, and N.~L. Johnson.
\newblock \emph{Continuous Multivariate Distributions. Volume 1: Models and
  Applications}.
\newblock Wiley, 2000.

\bibitem[Lindgren et~al.(2011)Lindgren, Rue, and Lindstr{\"o}m]{LindgRL2011_73}
F.~Lindgren, H.~Rue, and J.~Lindstr{\"o}m.
\newblock An explicit link between {G}aussian fields and {G}aussian {M}arkov
  random fields: the stochastic partial differential equation approach.
\newblock \emph{J. R. Stat. Soc. B}, 73\penalty0 (4):\penalty0 423--498, 2011.
\newblock \doi{10.1111/j.1467-9868.2011.00777.x}.

\bibitem[Makalic and Schmidt(2016)]{MakalS2016_23}
E.~Makalic and D.~F. Schmidt.
\newblock A simple sampler for the horseshoe estimator.
\newblock \emph{{IEEE} Signal Processing Lett.}, 23\penalty0 (1):\penalty0
  179--182, 2016.
\newblock \doi{10.1109/LSP.2015.2503725}.

\bibitem[Mikolajewicz et~al.(2007)Mikolajewicz, Gr{\"o}ger, Maier-Reimer,
  Schurgers, Vizca{\'\i}no, and Winguth]{MikolGMSVW2007_28}
U.~Mikolajewicz, M.~Gr{\"o}ger, E.~Maier-Reimer, G.~Schurgers,
  M.~Vizca{\'\i}no, and A.~M. Winguth.
\newblock Long-term effects of anthropogenic {CO}2 emissions simulated with a
  complex earth system model.
\newblock \emph{Clim. Dyn.}, 28\penalty0 (6):\penalty0 599--633, 2007.
\newblock \doi{10.1007/s00382-006-0204-y}.

\bibitem[Miller and Smith(2012)]{MilleS2012_41}
P.~A. Miller and B.~Smith.
\newblock Modelling tundra vegetation response to recent arctic warming.
\newblock \emph{Ambio}, 41\penalty0 (3):\penalty0 281--291, 2012.
\newblock \doi{10.1007/s13280-012-0306-1}.

\bibitem[Olofsson(2013)]{PhD_Olofsson2013}
J.~Olofsson.
\newblock \emph{The {Earth}: climate and anthropogenic interactions in a long
  time perspective}.
\newblock PhD thesis, Lund University, 2013.
\newblock URL \url{http://lup.lub.lu.se/record/3732052}.

\bibitem[Park and Casella(2008)]{ParkC2008_103}
T.~Park and G.~Casella.
\newblock The bayesian lasso.
\newblock \emph{J. Am. Stat. Assoc.}, 103\penalty0 (482):\penalty0 681--686,
  2008.
\newblock \doi{10.1198/016214508000000337}.

\bibitem[Pirzamanbein et~al.(2014)Pirzamanbein, Lindstr{\"o}m, Poska, Sugita,
  Trondman, Fyfe, Mazier, Nielsen, Kaplan, Bjune, Birks, Giesecke, Kangur,
  Lata{\l}owa, Marquer, Smith, and Gaillard]{PirzaLPSTFMNKBBGKLMSG2014_20}
B.~Pirzamanbein, J.~Lindstr{\"o}m, A.~Poska, S.~Sugita, A.-K. Trondman,
  R.~Fyfe, F.~Mazier, A.~Nielsen, J.~Kaplan, A.~Bjune, H.~Birks, T.~Giesecke,
  M.~Kangur, M.~Lata{\l}owa, L.~Marquer, B.~Smith, and M.-J. Gaillard.
\newblock Creating spatially continuous maps of past land cover from point
  estimates: A new statistical approach applied to pollen data.
\newblock \emph{Ecol. Complex.}, 20:\penalty0 127--141, 2014.
\newblock \doi{10.1016/j.ecocom.2014.09.005}.

\bibitem[Pirzamanbein et~al.(2018)Pirzamanbein, Lindstr{\"o}m, Poska, and
  Gaillard]{PirzaLPG2018_24}
B.~Pirzamanbein, J.~Lindstr{\"o}m, A.~Poska, and M.-J. Gaillard.
\newblock Modelling spatial compositional data: Reconstructions of past land
  cover and uncertainties.
\newblock \emph{Spatial Stat.}, 24:\penalty0 14--31, 2018.
\newblock \doi{10.1016/j.spasta.2018.03.005}.

\bibitem[Pitman et~al.(2009)Pitman, de~Noblet-Ducoudr{\'e}, Cruz, Davin, Bonan,
  Brovkin, Claussen, Delire, Ganzeveld, Gayler, van~den Hurk, Lawrence, van~der
  Molen, Müller, Reick, Seneviratne, Strengers, and
  Voldoire]{PitmaNCDBBCDGGHLMMRSSV2009_36}
A.~Pitman, N.~de~Noblet-Ducoudr{\'e}, F.~Cruz, E.~Davin, G.~Bonan, V.~Brovkin,
  M.~Claussen, C.~Delire, L.~Ganzeveld, V.~Gayler, B.~J. J.~M. van~den Hurk,
  P.~J. Lawrence, M.~K. van~der Molen, C.~Müller, C.~H. Reick, S.~I.
  Seneviratne, B.~J. Strengers, and A.~Voldoire.
\newblock Uncertainties in climate responses to past land cover change: First
  results from the {LUCID} intercomparison study.
\newblock \emph{Geophys. Res. Lett.}, 36\penalty0 (14):\penalty0 n/a--n/a,
  2009.
\newblock \doi{10.1029/2009GL039076}.

\bibitem[Päivinen et~al.(2001)Päivinen, Lehikoinen, Schuck, Häme, Väätäinen,
  Kennedy, and Folving]{PaeiviLSHVKF2001_Combining}
R.~Päivinen, M.~Lehikoinen, A.~Schuck, T.~Häme, S.~Väätäinen, P.~Kennedy, and
  S.~Folving.
\newblock Combining {Earth} observation data and forest statistics.
\newblock Technical Report~14, European Forest Institute, Joint Research
  Centre-European Commission., 2001.
\newblock URL
  \url{https://www.efi.int/publications-bank/combining-earth-observation-data-and-forest-statistics}.
\newblock ISBN: 952-9844-84-0 ISSN: 1238-8785.

\bibitem[Pongratz et~al.(2008)Pongratz, Reick, Raddatz, and
  Claussen]{PongrRRC2008_22}
J.~Pongratz, C.~Reick, T.~Raddatz, and M.~Claussen.
\newblock A reconstruction of global agricultural areas and land cover for the
  last millennium.
\newblock \emph{Glob. Biogeochem. Cycles}, 22\penalty0 (3):\penalty0 GB3018,
  2008.
\newblock \doi{10.1029/2007GB003153}.

\bibitem[Prentice et~al.(2007)Prentice, Bondeau, Cramer, Harrison, Hickler,
  Lucht, Sitch, Smith, and Sykes]{PrentBCHHLSSS2007_Canadell}
I.~C. Prentice, A.~Bondeau, W.~Cramer, S.~P. Harrison, T.~Hickler, W.~Lucht,
  S.~Sitch, B.~Smith, and M.~T. Sykes.
\newblock Dynamic global vegetation modeling: quantifying terrestrial ecosystem
  responses to large-scale environmental change.
\newblock In J.~G. Canadell, D.~E. Pataki, and L.~F. Pitelka, editors,
  \emph{Terrestrial Ecosystems in a Changing World. Global Change --- The IGBP
  Series}, pages 175--192. Springer, 2007.
\newblock \doi{10.1007/978-3-540-32730-1_15}.

\bibitem[Richter-Menge et~al.(2011)Richter-Menge, Jeffries, and
  Overland]{RichtJO2011}
J.~A. Richter-Menge, M.~O. Jeffries, and J.~E. Overland, editors.
\newblock \emph{Arctic Report Card 2011}.
\newblock National Oceanic and Atmospheric Administration, 2011.
\newblock URL \url{www.arctic.noaa.gov/reportcard}.

\bibitem[Rue and Held(2005)]{RueH2005}
H.~Rue and L.~Held.
\newblock \emph{{Gaussian} {Markov} Random Fields; Theory and Applications},
  volume 104 of \emph{Monographs on Statistics and Applied Probability}.
\newblock Chapman \& Hall/CRC, 2005.

\bibitem[Samuelsson et~al.(2011)Samuelsson, Jones, Will{\'e}n, Ullerstig,
  Gollvik, Hansson, Jansson, Kjellstr{\"o}m, Nikulin, and
  Wyser]{SamueJWUGHJKNW2011_63}
P.~Samuelsson, C.~G. Jones, U.~Will{\'e}n, A.~Ullerstig, S.~Gollvik,
  U.~Hansson, C.~Jansson, E.~Kjellstr{\"o}m, G.~Nikulin, and K.~Wyser.
\newblock The {Rossby} {Centre} regional climate model {RCA3}: model
  description and performance.
\newblock \emph{Tellus A}, 63\penalty0 (1):\penalty0 4--23, 2011.

\bibitem[Scheiter et~al.(2013)Scheiter, Langan, and Higgins]{ScheiLH2013_198}
S.~Scheiter, L.~Langan, and S.~I. Higgins.
\newblock Next-generation dynamic global vegetation models: learning from
  community ecology.
\newblock \emph{New Phytologist}, 198\penalty0 (3):\penalty0 957--969, 2013.
\newblock \doi{10.1111/nph.12210}.

\bibitem[Schuck et~al.(2002)Schuck, van Brusselen, P{\"a}ivinen, H{\"a}me,
  Kennedy, and Folving]{SchucBPHKF2002_Compilation}
A.~Schuck, J.~van Brusselen, R.~P{\"a}ivinen, T.~H{\"a}me, P.~Kennedy, and
  S.~Folving.
\newblock Compilation of a calibrated {European} forest map derived from
  {NOAA}-{AVHRR} data.
\newblock EFI Internal Report~13, EuroForIns, 2002.

\bibitem[Sitch et~al.(2003)Sitch, Smith, Prentice, Arneth, Bondeau, Cramer,
  Kaplan, Levis, Lucht, Sykes, Thonicke, and Venevsky]{SitchSPABCKLLSTV2003_9}
S.~Sitch, B.~Smith, I.~C. Prentice, A.~Arneth, A.~Bondeau, W.~Cramer,
  J.~Kaplan, S.~Levis, W.~Lucht, M.~Sykes, K.~Thonicke, and S.~Venevsky.
\newblock Evaluation of ecosystem dynamics, plant geography and terrestrial
  carbon cycling in the {LPJ} dynamic global vegetation model.
\newblock \emph{Glob. Change Biol.}, 9\penalty0 (2):\penalty0 161--185, 2003.

\bibitem[Smith et~al.(2001)Smith, Prentice, and Sykes]{SmithPS2001_10}
B.~Smith, I.~C. Prentice, and M.~T. Sykes.
\newblock Representation of vegetation dynamics in the modelling of terrestrial
  ecosystems: comparing two contrasting approaches within {European} climate
  space.
\newblock \emph{Glob. Ecol. Biogeogr.}, 10:\penalty0 621--637, 2001.

\bibitem[Strandberg et~al.(2011)Strandberg, Brandefelt, Kjellstr{\"o}m, and
  Smith]{StranBKS2011_63}
G.~Strandberg, J.~Brandefelt, E.~Kjellstr{\"o}m, and B.~Smith.
\newblock High-resolution regional simulation of last glacial maximum climate
  in {Europe}.
\newblock \emph{Tellus A}, 63\penalty0 (1):\penalty0 107--125, 2011.

\bibitem[Strandberg et~al.(2014)Strandberg, Kjellstr\"om, Poska, Wagner,
  Gaillard, Trondman, Mauri, Davis, Kaplan, Birks, Bjune, Fyfe, Giesecke,
  Kalnina, Kangur, van~der Knaap, Kokfelt, Kune\v{s}, Lata\l~owa, Marquer,
  Mazier, Nielsen, Smith, Sepp\"a, and
  Sugita]{StranKPWGTMDKBBFGKKKKKLMMNSSS2014_10}
G.~Strandberg, E.~Kjellstr\"om, A.~Poska, S.~Wagner, M.-J. Gaillard, A.-K.
  Trondman, A.~Mauri, B.~A.~S. Davis, J.~O. Kaplan, H.~J.~B. Birks, A.~E.
  Bjune, R.~Fyfe, T.~Giesecke, L.~Kalnina, M.~Kangur, W.~O. van~der Knaap,
  U.~Kokfelt, P.~Kune\v{s}, M.~Lata\l~owa, L.~Marquer, F.~Mazier, A.~B.
  Nielsen, B.~Smith, H.~Sepp\"a, and S.~Sugita.
\newblock Regional climate model simulations for europe at 6 and 0.2 k bp:
  sensitivity to changes in anthropogenic deforestation.
\newblock \emph{Clim. Past}, 10\penalty0 (2):\penalty0 661--680, 2014.
\newblock \doi{10.5194/cp-10-661-2014}.
\newblock URL \url{http://www.clim-past.net/10/661/2014/}.

\bibitem[Trondman et~al.(2015)Trondman, Gaillard, Sugita, Mazier, Fyfe,
  Lechterbeck, Marquer, Nielsen, Twiddle, Barratt, Birks, Bjune, Caseldine,
  David, Dodson, D{\"o}rfler, Fischer, Giesecke, Hultberg, Kangur, Kune\v{s},
  Lata{\l}owa, Leydet, Lindbaldh, Mitchell, Odgaard, Peglar, Persson,
  R{\"o}sch, van~der Knaap, van Geel, Smith, and
  Wick]{TrondGSMFLMNTBBBCDDDFGHKKLLLMOPPRKGSW2015_21}
A.-K. Trondman, M.-J. Gaillard, S.~Sugita, F.~Mazier, R.~Fyfe, J.~Lechterbeck,
  L.~Marquer, A.~Nielsen, C.~Twiddle, P.~Barratt, H.~Birks, A.~Bjune,
  C.~Caseldine, R.~David, J.~Dodson, W.~D{\"o}rfler, E.~Fischer, T.~Giesecke,
  T.~Hultberg, M.~Kangur, P.~Kune\v{s}, M.~Lata{\l}owa, M.~Leydet,
  M.~Lindbaldh, F.~Mitchell, B.~Odgaard, S.~Peglar, T.~Persson, M.~R{\"o}sch,
  P.~van~der Knaap, B.~van Geel, A.~Smith, and L.~Wick.
\newblock Pollen-based quantitative reconstructions of past land-cover in {NW
  Europe} between 6k years {BP} and present for climate modelling.
\newblock \emph{Glob. Change Biol.}, 21\penalty0 (2):\penalty0 676--697, 2015.
\newblock \doi{10.1111/gcb.12737}.

\bibitem[Zhang et~al.(2013)Zhang, Miller, Smith, Wania, Koenigk, and
  D{\"o}scher]{ZhangMSWKD2013_8}
W.~Zhang, P.~A. Miller, B.~Smith, R.~Wania, T.~Koenigk, and R.~D{\"o}scher.
\newblock Tundra shrubification and tree-line advance amplify arctic climate
  warming: results from an individual-based dynamic vegetation model.
\newblock \emph{Environ. Res. Lett.}, 8\penalty0 (3):\penalty0 034023, 2013.

\end{thebibliography}

\end{document}